# 14
# Could Inelastic Interactions Induce Quantum Probabilistic Transitions?

Nicholas Maxwell

What are quantum entities? Is the quantum domain deterministic or probabilistic? Orthodox quantum theory (OQT) fails to answer these two fundamental questions. As a result of failing to answer the first question, OQT is very seriously defective: it is imprecise, ambiguous, ad hoc, non-explanatory, inapplicable to the early universe, inapplicable to the cosmos as a whole, and such that it is inherently incapable of being unified with general relativity. It is argued that probabilism provides a very natural solution to the quantum wave/particle dilemma and promises to lead to a fully micro-realistic, testable version of quantum theory that is free of the defects of OQT. It is suggested that inelastic interactions may induce quantum probabilistic transitions.

## 14.1 Two Fundamental Questions

What are quantum entities in view of their baffling wave and particle properties? Is the quantum domain deterministic or probabilistic? Even though quantum theory was invented more than 90 years ago, there are still no agreed answers to these two simple and fundamental questions.

In a series of papers published some years ago,[1] I have argued that the answer to the second question is that the quantum domain is fundamentally probabilistic, this answer having the great virtue that it provides a very natural solution to the first question—the baffling quantum wave/particle dilemma.[2] A few words of explanation.

We need to appreciate that there is a rough-and-ready one-to-one correspondence between physical properties of entities on the one hand and dynamical laws of the relevant theory on the other hand.[3] Thus, in speaking of the physical properties of physical entities—electric charge, mass, wavelength, and so on—we are, in effect, speaking of dynamic laws that govern the way these entities evolve and interact. And vice versa: in speaking of dynamic laws, we are, in effect, speaking of physical properties that entities, postulated by the theory, will possess. Thus, if we change, more or less radically, the *laws* that govern the behaviour of entities, we thereby change, more or less radically, the *physical properties* of these entities, the kind of physical entities that they are.

The classical particle and the classical wave (or field) belong to classical, deterministic physics. If, now, we move from the deterministic laws of classical physics to fundamentally probabilistic laws of quantum theory, we must thereby change radically the physical nature of the entities that probabilistic quantum theory would be about.

Granted that quantum theory is indeed fundamentally probabilistic, so that the physical entities of the quantum domain interact with one another *probabilistically*, then the entities of the quantum domain must be radically different from anything encountered within deterministic, classical physics. It would indeed be a disaster, as far as the comprehensibility of the quantum domain is concerned—assuming it is fundamentally probabilistic—if quantum entities turned out to be classical particles or classical waves (or fields), as both are to be associated with *determinism*. That quantum entities—electrons, photons, atoms, and the rest—turn out to be very different from deterministic classical particles or waves is thus excellent news from the standpoint of the intelligibility of the quantum domain.

---

[1] See Maxwell (1972; 1976; 1982; 1988; 1994; 1998, ch. 7; 2004; 2011). See also Maxwell (2017b, pp. 135–151).
[2] At the time of my first publication on fully micro-realistic, fundamentally probabilistic quantum theory, in 1972, there were very few others working on this approach. Since then, theoretical and experimental work on probabilistic collapse has become a growing field of research. For a recent survey of approaches and literature, see Bassi et al. (2013).
[3] This point is spelled out in much greater detail in Maxwell (1976; 1988); see also Maxwell (1968; 1998, pp. 141–155; 2017a, pp. 116–118).

The founding fathers of quantum theory struggled to find an answer to the question "Are quantum entities particles or waves?" In the end, most physicists despaired of finding an answer, came to the conclusion that the question of what sort of entities quantum entities are could not be answered, and adopted the orthodox interpretation of quantum theory, which evades the whole issue by restricting quantum theory to being a theory about the results of performing *measurements*.[4]

What everyone failed to appreciate—both those who accepted orthodox quantum theory (OQT) following Bohr and Heisenberg and those who were critical of OQT following Einstein and Schrödinger—is that the key question that most physicists despaired of answering, namely "Are quantum entities particles or waves?" is the *wrong* question to ask. The *right* questions, granted the crucial assumption that the quantum domain is fundamentally probabilistic, are:

(1) What sort of *unproblematic*, fundamentally probabilistic entities are there as possibilities?

(2) Are quantum entities some variety of a kind of unproblematic, fundamentally probabilistic entity?

In short, what everyone still tends to take for granted—namely that the quantum domain is inherently baffling and incomprehensible because it cannot be made sense of in terms of the classical particle or the classical wave (or field)—is actually very good news indeed as far as the intelligibility of the quantum domain is concerned. The quantum domain would have been incomprehensible indeed if composed of classical, deterministic particles or waves. That it is *not* composed of such entities is very encouraging from the standpoint of the quantum domain being thoroughly intelligible.[5]

**14.2 Is the Quantum Domain Fundamentally Probabilistic?**

The foregoing considerations depend, crucially, on the quantum domain being fundamentally probabilistic. What grounds do we have for holding this to be true?

There are two possibilities: determinism and probabilism. Both should be explored.[6] Here, I explore some consequences of adopting the probabilistic hypothesis. Not only does this hypothesis imply that the traditional "Wave or particle?" question is the wrong question to ask. It has, as I hope to make clear in a moment, the great virtue of rendering the behaviour and properties of quantum entities thoroughly comprehensible in an entirely natural way.

---

[4] This was the situation up to around 1990 perhaps. Increasingly, in the last two decades or so, physicists have come to appreciate that the orthodox interpretation of quantum theory is unsatisfactory, even unacceptable. It is now acceptable for physicists to publish research on the subject. This is strikingly borne out by the present volume and by Shan Gao's *Meaning of the Wave Function* (2017), which provides a sustained exploration of interpretative problems of quantum theory and gives an indication of the vast and growing literature on the subject. For a recent survey of fundamentally probabilistic or "collapse" versions of quantum theory, see Bassi et al. (2013).

[5] For a brief discussion of the point that, though hints of probabilism can be found in the development of quantum theory from the outset, nevertheless probabilistic quantum realism was long ignored, see Maxwell (2011, sections 12 and 13).

[6] At the time of writing (2016), the most popular deterministic version of quantum theory seems to be the Everett or many-worlds interpretation: for a recent exposition see Wallace (2012); for discussion see Bacciagaluppi and Ismae (2015). For my criticisms see Maxwell (2017b, pp. 148–151).

Probabilism has additional virtues. It must lead, relentlessly, to a version of quantum theory that makes some empirical predictions that differ from those of OQT. It leads to testable physics, in other words.

The point is this. OQT, in general, makes probabilistic predictions about the results of measurement. Measurement is, thus, a *sufficient* condition for probabilistic events to occur (granted that we are not interpreting such measurements in deterministic terms). What is not acceptable, however, is that measurement should be a *necessary* condition for probabilistic events. If nature is fundamentally probabilistic, then probabilistic events will occur whether or not there are physicists around to make measurements. They will have occurred long before life began on earth. Probabilistic quantum theory (PQT) is thus committed to holding that there must be fully *quantum mechanical* physical conditions for probabilistic transitions to occur, conditions which will obtain in nature entirely in the absence of measurement. Any version of PQT which specifies precisely what these conditions are must make predictions that differ from those of OQT (which restricts probabilism to measurement). The major task confronting the development of PQT is to specify precisely what the physical conditions for probabilistic events to occur are.

PQT has further potential advantages. It leads to a fully micro-realistic theory, one which specifies how quantum entities evolve and interact physically, in physical space and time, entirely independently of preparation and measurement. It is thus able to explain how the classical world emerges, as an approximation, from the quantum world. OQT, because it fails to solve the wave/particle problem and thus fails to be a fully micro-realistic theory about quantum entities (there being no coherent idea as to what such entities are) must, as a result, be a theory about the results of measurement. This means some part of classical physics (or some other theory) must be used to describe the measuring instrument. This in turn means that OQT suffers from the very serious defects that it is imprecise, ambiguous, ad hoc, non-explanatory, inapplicable to the early universe, inapplicable to the cosmos as a whole, and such that it is inherently incapable of being unified with general relativity.[7] PQT is free of these seven very serious defects precisely because it has its own micro-quantum ontology (it is about "beables" as John Bell called them), and thus makes no reference to measurement in its basic postulates. All these seven defects of OQT arise because the basic postulates of OQT refer to observables and thus to measurement. And PQT is free of the key eighth defect of OQT, responsible for all the others, namely the failure of OQT to provide a coherent micro-ontology for the quantum domain.

There are, in short, decisive grounds for developing a better version of quantum theory than OQT, whether deterministic or probabilistic.

### 14.3 What Kinds of Possible Fundamentally Probabilistic Entity Are There?

How, in general terms, is the fundamentally probabilistic physical entity to be understood? Just as probabilism generalizes determinism, so too the idea of the fundamentally probabilistic object generalizes the notion of the deterministic object, familiar from common sense and classical physics. Objects of common sense and of classical deterministic physics have dispositional or necessitating properties which determine how the object in question behaves in certain circumstances.[8] A ball that is *elastic* bounces when dropped onto the floor. An *inflammable* object catches fire when exposed to a naked flame. A body with *Newtonian gravitational charge* obeys Newton's law $F = Gm_1m_2/d^2$. Fundamentally probabilistic objects, which interact with one another probabilistically, and which I shall call *propensitons*, have physical properties which generalize the familiar, deterministic notion of physical property, and which determine how objects behave *probabilistically* in appropriate circumstances. An example of such a probabilistic property, or *propensity*,[9] is what may be called the "bias" of a die—the property of the die which determines the probabilities of the outcomes 1 to 6 when the die is tossed onto a table. A value of bias assigns a probability to each of the six possible outcomes. We can even imagine that the value of the bias of the die itself changes: there is, perhaps, a tiny magnet imbedded in the die and an electromagnet under the table. As the strength of the magnet beneath the table varies, so the value of the bias of the die will change.

---

[7] These points are argued for in some detail in Maxwell (1988, pp. 1–8; 1993, section iv; 2004, section 1). Some of these points have been argued for even earlier, in Maxwell (1972; 1976; 1982). Bell (1973; 1987) also argued that the concept of measurement needs to be eliminated from the basic postulates of quantum theory, the theory being interpreted to be about "beables" rather than "observables." Bell tended to restrict himself, however, to the point that, if quantum theory is about measurement, it is inherently *imprecise*. I discuss Bell's contribution in Maxwell (1992).

The idea, then, is to interpret the ψ-function of quantum theory as attributing values of physical *propensities* to quantum propensitons, thus specifying the real physical states of propensitons in physical space and time. The ψ-function carries implications about how quantum propensitons interact with one another probabilistically should appropriate quantum conditions for such interactions to occur arise.

Granted we adopt probabilism as our working hypothesis, the crucial question to ask is: What different kinds of propensiton are there as possibilities? There are two cases to consider. Probabilistic events might occur (i) continuously in time or (ii) discontinuously, only when specific physical conditions arise.

The structure of OQT suggests that we should initially favour (ii). Given the astonishing empirical success of OQT—the wealth and variety of phenomena it predicts, the accuracy of its predictions, and the lack of empirical refutations—it is clear that OQT has got quite a lot right. All the defects of OQT stem from the failure to solve the wave/particle problem and the resulting need to formulate the theory as being about measurement. This suggests that we should, initially at least, keep as close to OQT as possible and modify the theory just sufficiently to remove these defects. According to OQT, probabilistic events only occur when measurements are made; quantum states evolve deterministically, in accordance with the Schrödinger equation, when no measurement is made. In developing PQT, then, we should, initially, favour (ii), hold onto the Schrödinger equation when no probabilistic events occur, and replace measurement with some physical condition more plausible to be the condition for the occurrence of probabilistic events. Let us call a physical entity that evolves deterministically until precise physical conditions arise for probabilistic transitions to occur an "intermittent propensiton."

Here is a very simple-minded model of a world made up of such intermittent propensitons. The world is composed of spheres, which expand deterministically at a steady rate. The instant two spheres touch, they collapse to a pre-determined minute sphere, each within its spherical volume, the precise location only probabilistically determined. This, one might say, is the simplest, the most elementary, and the most comprehensible example of the intermittent propensiton.

---

[8] For this account of "necessitating" physical properties, see Maxwell (1968; 1998, pp. 141–155).
[9] Popper introduced the idea of propensities in connection with interpretative problems of QT; see Popper (1957; 1967; 1982), although, as Popper (1982, pp. 130–135) has pointed out, Born, Heisenberg, Dirac, Jeans, and Landé have all made remarks in this direction. The version of the propensity idea employed here is, however, in a number of respects, different from and an improvement over the notion introduced by Popper: see Maxwell (1976, pp. 284–286; 1985, pp. 41–42). For a discussion of Popper's contributions to the interpretative problems of quantum theory, see Maxwell (2016, section 6, especially note 19).

A slightly more complicated model is the following. Each deterministically expanding sphere is filled with a "stuff," which varies in density, from place to place. Where the "stuff" is dense, it is correspondingly more probable that it is there that the tiny sphere will be located when the probabilistic collapse occurs; where the "stuff" is sparse, it is correspondingly less probable.

We can now imagine five further modifications. First, the "stuff"—which we may call "position probability density"—may vary in a wavelike fashion within the sphere. Second, the propensiton ceases to be a sphere and may acquire any shape and size whatsoever within physical space. Third, we suppose that the physical state of the propensiton, from moment to moment in physical space, is specified by a complex function $\psi(r,t)$. Fourth, the evolution of the propensiton state is determined by Schrödinger's time-dependent equation. And fifth, $|\psi|^2 dv$ specifies the density of the "stuff" of the propensiton in volume element $dv$—that is, the probability of the propensiton undergoing a probabilistic interaction in $dv$ should the physical conditions for such an interaction to occur obtain.

We began with the simplest, the most elementary, the most intelligible example of an intermittent propensiton conceivable. We have added a few details, and we have come up with a physical entity which behaves in space and time almost precisely in accordance with quantum theory. The quantum intermittent propensiton may seem mysterious, but that is just because ordinarily we have no experience of such entities. The quantum propensiton just indicated is in fact a special case of the simplest, the most intelligible intermittent propensiton conceivable, in turn a probabilistic generalization of the ordinary, quasi-deterministic objects we are familiar with: footballs, tables, pencils.

In the next section, I reply to objections to my claim that the foregoing provides us with a fully intelligible, micro-realistic version of quantum theory. In the section after that, I put forward my proposal about what it is, physically, that induces probabilistic events to occur.

### 14.4 Can the $\psi$-function be Interpreted to Specify the Physical State of the Propensiton in Physical Space?

**Objection (1):** The $\psi$-function is complex and hence cannot be employed to describe the physical state of an actual physical system.

**Reply:** The complex $\psi$ is equivalent to two interlinked real functions, which can be regarded as specifying the propensity state of quantum systems. In any case, as Penrose (2004, p. 539) reminds us, complex numbers are used in classical physics, without this creating a problem concerning the reality of what is described. The complex nature of $\psi$ has to do, in part, with the fact that the wave-like character of $\psi$ is not in physical space, except when interference leads to spatio-temporal wave-like variations in the intensity of $\psi$ and thus in $|\psi(x,t)|^2$ as well.

**Objection (2):** Given a physical system of N quantum entangled systems, the $\psi$-function is a function of 3N-dimensional configuration space and not 3-dimensional physical space. This makes it impossible to interpret such a $\psi$-function as specifying the physical state of N quantum entangled physical systems in physical space when $N \geq 2$.

**Reply**: $\psi(r_1, r_2 \ldots r_N)$ can be regarded as assigning a complex number to any point in 3N-dimensional configuration space. Equally, however, we can regard $\psi(r_1, r_2 \ldots r_N)$ as assigning the complex number to N points in 3-dimensional physical space. Suppose $\psi(r_1, r_2 \ldots r_N)$ is the quantum entangled state of N distinct kinds of particle. Then $\psi(r_1, r_2 \ldots r_N)$, in assigning a complex number to a point in configuration space, can be interpreted as assigning this number to N points in physical space, each point labelled by a different particle. The quantum propensiton state in physical space will be multi-valued at any point in physical space and also highly non-local, in that its values at any given point cannot be dissociated from values at N-1 other points. If we pick out N distinct points in physical space, there will be, in general, N! points in configuration space which assign different values of ψ to these N physical points, corresponding to the different ways the N particles can be reassigned to these N points. If we pick out just one point in physical space $(x_o,y_o,z_o)$, the ψ-function will in general assign infinitely many different complex numbers to this point $(x_o,y_o,z_o)$, corresponding to different locations of the particles in physical space—there being infinitely many points in configuration space that assign a complex number to this point $(x_o,y_o,z_o)$ in physical space. The N-particle, quantum-entangled propensiton is, in physical space, a complicated, non-local, multi-valued object, very different from anything found in classical physics. Its physical nature in 3-dimensional physical space is, nevertheless, precisely specified by the single-valued $ψ(r_1,r_2 \ldots r_N)$ in 3N-dimensional configuration space.[10]

**Objection (3)**: The ψ-function is highly non-local in character. This, again, makes a realistic interpretation of it impossible.

**Reply**: As my reply to objection (2) indicates, quantum propensitons of the type being considered here, made up of a number of quantum entangled "particles," are highly non-local in character, in that one cannot specify what exists at one small region of physical space without simultaneously taking into account what exists at other small regions. Propensitons of this type seem strange because they are unfamiliar—but we must not confuse the unfamiliar with the inexplicable or impossible. Non-local features of the ψ-function do not prevent it from specifying the actual physical states of propensitons; propensitons just are, according to the version of PQT being developed here, highly non-local objects in the sense indicated.

**Objection (4)**: If the ψ-function is interpreted realistically, it follows that when a position measurement is made, a quantum system that has a state spread throughout a large volume of space collapses instantaneously into a minute region where the system is detected. Such an instantaneous collapse, possibly across vast distances of space, becomes wholly implausible when regarded as a real physical process.

**Reply**: Such instantaneous probabilistic collapse is an inherent feature of the intermittent propensiton. There is nothing implausible or inexplicable about such probabilistic transitions.[11] To suppose otherwise is to be a victim of deterministic prejudice—a victim of the dogma that nature must be deterministic, the more general idea of probabilism being excluded a priori, without any valid reason.

I conclude that there are no valid objections to interpreting ψ as specifying the actual physical states of propensitons in physical space.

### 14.5 Do Inelastic Interactions Induce Probabilistic Collapse?

In order to specify the precise nature of the quantum intermittent propensitons under consideration and at the same time give precision to the version of PQT being developed here, we need now to specify (a) the precise quantum conditions for a probabilistic transition to occur in a quantum system, (b) what the possible outcome quantum states are, given that the quantum state at the instant of probabilistic transition is ψ, and (c) how ψ assigns probabilities to the possible outcomes. No reference must be made to measurement, observables, macroscopic systems, irreversible processes, or classically described systems.

---

[10] This solution to the problem was outlined in Maxwell (1976, pp. 666–667; and 1982, p. 610). Albert (1996) has proposed that the quantum state of an N-particle entangled system be interpreted to exist physically in 3N-dimensional configuration space. But configuration space is a mathematical fiction, not a physically real arena in which events occur. Albert's proposal is untenable, and in any case unnecessary.

[11] Instantaneous probabilistic collapse is, however, highly problematic the moment one considers developing a Lorentz-invariant version of the theory. For my views about this problem, see Maxwell (1985; 2006; 2011; 2017c). For a fascinating discussion of the problem of reconciling special relativity and the non-locality of quantum theory, see Maudlin (2011).

One possibility is the proposal of Ghirardi, Rimini, and Weber (1986)—see also Ghirardi (2002)—according to which the quantum state of a system such as an electron collapses spontaneously, on average after the passage of a long period of time, into a highly localized state. When a measurement is performed on the electron, it becomes quantum entangled with millions upon millions of quantum systems that go to make up the measuring apparatus. In a very short time, there is a high probability that one of these quantum systems will spontaneously collapse, causing all the other quantum-entangled systems, including the electron, to collapse as well. At the micro level, it is almost impossible to detect collapse, but at the macro level, associated with measurement, collapse occurs very rapidly all the time.

Another possibility is the proposal of Penrose (1986, 2004, ch. 30), according to which collapse occurs when the state of a system evolves into a superposition of two or more states, each state having, associated with it, a sufficiently large mass located at a distinct region of space. The idea is that general relativity imposes a restriction on the extent to which such superpositions can develop, in that it does not permit such superpositions to evolve to such an extent that each state of the superposition has a substantially distinct space–time curvature associated with it.

The possibility that I favour, put forward before either Ghirardi, Rimini, and Weber's proposal or Penrose's proposal, is that probabilistic transitions occur whenever, as a result of inelastic interactions between quantum systems, new "particles," new bound, stationary, or decaying systems, are created (Maxwell, 1972, 1976, 1982, 1988, 1994). A little more precisely:

(I) Whenever, as a result of an inelastic interaction, a system of interacting "particles" creates new "particles," bound, stationary, or decaying systems, so that the state of the system goes into a superposition of states, each state having associated with it different particles or bound, stationary, or decaying systems, then, when the interaction is nearly at an end, spontaneously and probabilistically, entirely in the absence of measurement, the superposition collapses into one or other state.

An example of the kind of inelastic interaction that is involved is the following:

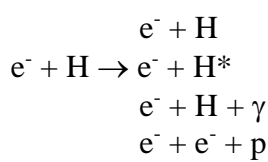

Here $e^-$, H, H*, γ, and p stand for electron, hydrogen atom, excited hydrogen atom, photon, and proton, respectively.)

What exactly does it mean to assert that the "interaction is very nearly at an end" in the previous postulate? My suggestion, here, is that it means that forces between the "particles" are very nearly zero except for forces holding bound or decaying systems together. In order to indicate how this can be formulated precisely, consider the toy interaction:

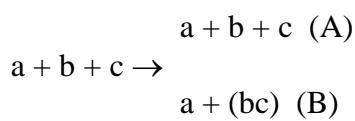

Here, a, b, and c are spinless particles, and (bc) is the bound system. Let the state of the entire system be Φ(t), and let the asymptotic states of the two channels (A) and (B) be ψ$_A$(t) and ψ$_B$(t) respectively. Asymptotic states associated with inelastic interactions are fictional states towards which, according to OQT, the real state of the system evolves as t → + ∞. Each outcome channel has its associated asymptotic state, which evolves as if forces between particles are zero, except where forces hold bound systems together.

According to OQT, in connection with the toy interaction, there are states φ$_A$(t) and φ$_B$(t), such that:

(1) For all t, Φ(t) = c$_A$φ$_A$(t) + c$_B$φ$_B$(t), with |c$_A$|$^2$ + |c$_B$|$^2$ = 1;
(2) as t → + ∞, φ$_A$(t) → ψ$_A$(t) and φ$_B$(t) → ψ$_B$(t).

According to the version of PQT under consideration here, at the first instant t for which φ$_A$(t) is very nearly the same as the asymptotic state ψ$_A$(t), or φ$_B$(t) is very nearly the same as ψ$_B$(t), then the state of the system, Φ(t), collapses spontaneously either into φ$_A$(t) with probability |c$_A$|$^2$ or into φ$_B$(t) with probability |c$_B$|$^2$. Or, more precisely:

(II) At the first instant for which | ⟨ψ$_A$(t)|φ$_A$(t)⟩ |$^2$ > 1 — ε or | ⟨ψ$_B$(t)|φ$_B$(t)⟩ |$^2$ > 1 — ε, the state of the system collapses spontaneously into φ$_A$(t) with probability |c$_A$|$^2$ or into φ$_B$(t) with probability |c$_B$|$^2$, ε being a universal constant, a positive real number very nearly equal to zero.[12]

According to (II), if ε = 0, probabilistic collapse occurs only when t = + ∞ and the corresponding version of PQT becomes equivalent to the many-worlds, or Everett, interpretation of quantum theory. As ε is chosen to be closer and closer to 1, so collapse occurs more and more rapidly, for smaller and smaller times t—and the corresponding versions of PQT become more and more readily falsifiable experimentally.

The evolutions of the actual state of the system, Φ(t), and the asymptotic states, ψ$_A$(t) and ψ$_B$(t), are governed by the respective channel Hamiltonians, H, H$_A$, and H$_B$, where:

$$H = - \left( \frac{\hbar^2 \nabla_a^2}{2m_a} + \frac{\hbar^2 \nabla_b^2}{2m_b} + \frac{\hbar^2 \nabla_c^2}{2m_c} \right) + V_{ab} + V_{bc} + V_{ac}$$

$$H_A = - \left( \frac{\hbar^2 \nabla_a^2}{2m_a} + \frac{\hbar^2 \nabla_b^2}{2m_b} + \frac{\hbar^2 \nabla_c^2}{2m_c} \right)$$

$$H_B = = - \left( \frac{\hbar^2 \nabla_a^2}{2m_a} + \frac{\hbar^2 \nabla_b^2}{2m_b} + \frac{\hbar^2 \nabla_c^2}{2m_c} \right) + V_{bc}$$

Here, m$_a$, m$_b$, and m$_c$ are the masses of "particles" a, b, and c, respectively, and ℏ = h/2π where h is Planck's constant.

This condition for probabilistic collapse can readily be generalized to apply to more complicated and realistic inelastic interactions between quantum systems.

According to this fully micro-realistic, fundamentally probabilistic version of quantum theory, the state function, Φ(t), describes the actual physical state of the quantum system—the propensiton—from moment to moment. The physical (quantum) state of the propensiton evolves in accordance with Schrödinger's time-dependent equation as long as the condition for a probabilistic transition to occur does not arise. The moment it does arise, the state jumps instantaneously and probabilistically, in the manner indicated, into a new state. (All but one of a superposition of states, each with distinct "particles" associated with them, vanish.) The new state then continues to evolve in accordance with Schrödinger's equation until conditions for a new probabilistic transition arise. Quasi-classical objects arise as a result of the occurrence of a rapid sequence of many such probabilistic transitions.

---

[12] The basic idea of (II) is to be found in Maxwell (1982 and 1988). It was first formulated precisely in Maxwell (1994).

### 14.6 Propensiton PQT Recovers All the Empirical Success of OQT

The version of propensiton quantum theory (PQT) just indicated recovers—in principle—all the empirical success of orthodox quantum theory (OQT). In order to see this, it is crucial to take note of the distinction between *preparation* and *measurement* (Popper, 1959, pp. 225–226; Margenau, 1958, 1963). A preparation is some physical procedure which has the consequence that if a quantum system exists (or is found) in some predetermined region of space, then it will have (or will have had) a definite quantum state. A measurement, by contrast, actually detects a quantum system and does so in such a way that a value can be assigned to some quantum "observable" (position, momentum, energy, spin, etc.). A measurement need not be a preparation. Measurements of photons, for example, far from preparing the photons to be in some quantum state, usually *destroy* the photons measured! On the other hand, a preparation is not in itself a measurement, because it does not *detect* what is prepared. It can be converted into a measurement by a subsequent detection.

From the formalism of OQT, one might well suppose that the various quantum observables are all on the same level and have equal status. This is in fact not the case. Position is fundamental,[13] and measurements of all other observables are made up of a combination of preparations and position measurements.[14] PQT, in order to do justice to quantum measurements, need only do justice to *position* measurements.

It might seem, to begin with, that PQT, based on the two postulates (I) and (II), which say nothing about position or localization, cannot predict that unlocalized systems become localized, necessary, it would seem, to predict the outcome of position measurements. The version of propensiton PQT just formulated does, however, predict that localizations occur. If a highly localized system, $S_1$, interacts inelastically with a highly unlocalized system, $S_2$, in such a way that a probabilistic transition occurs, then $S_1$ will localize $S_2$. If an atom or nucleus emits a photon or other "particle" which travels outwards in a spherical shell and which is subsequently absorbed by a localized third system, the localization of the photon or "particle" will localize the emitting atom or nucleus with which it was quantum entangled.[15]

---

[13] The possibility of formulating quantum theory so that it makes predictions exclusively about positions is made by Feynman and Hibbs, who write, "all measurements of quantum mechanical systems could be made to reduce eventually to position and time measurements..." Because of this possibility a theory formulated in terms of position measurements is complete enough in principle to describe all phenomena: see Feynman and Hibbs (1965, p. 96).

[14] Popper distinguished preparation and measurement in part in order to make clear that Heisenberg's uncertainty relations prohibit the simultaneous *preparation* of systems in a precise state of position and momentum but place no restrictions whatsoever on the simultaneous *measurement* of position and measurement. One needs, indeed, to measure position and momentum simultaneously well within the Heisenberg uncertainty relations simply to check up experimentally on the predictions of these relations: see Popper (1959, pp. 223–236). See Maxwell (2016, section 6) for a discussion of the important contribution that Popper makes, here, to the interpretation of orthodox quantum theory.

[15] If postulates (I) and (II) procure insufficient localization, (II) could be modified so that the outcome states are those that would have evolved from a spatial region characteristic of the size of the "particles" involved or characteristic of the distance associated with the inelastic interaction. Postulate

That PQT recovers (in principle) all the empirical success of OQT is a consequence of the following four points.[16]

First, OQT and PQT use the same dynamical equation, namely Schrödinger's time-dependent equation.

Second, whenever a position measurement is made and a quantum system is detected, this invariably involves an inelastic interaction and the creation of a new "particle" (bound or stationary system, such as the ionisation of an atom or the dissociation of a molecule, usually millions of these). This means that whenever a position measurement is made, the conditions for probabilistic transitions to occur, according to PQT, are satisfied. PQT will reproduce the predictions of OQT (given that PQT is provided with a specification of the quantum state of the measuring apparatus). As an example of PQT predicting, probabilistically, the result of a position measurement, consider the following. An electron in the form of a spatially spread-out wave packet is directed towards a photographic plate. According to PQT, the electron wave packet (or propensiton) interacts with billions of silver bromide molecules spread over the photographic plate: these evolve momentarily into superpositions of the dissociated and undissociated states until the condition for probabilistic collapse occurs, and just one silver bromide molecule is dissociated, and all the others remain undissociated. When the plate is developed (a process which merely makes the completed position measurement more visible), it will be discovered that the electron has been detected as a dot in the photographic plate.

Third, all other observables of OQT, such as momentum, energy, angular momentum, or spin, always involve (i) a preparation procedure which leads to distinct spatial locations being associated with distinct values of the observable to be measured and (ii) a position measurement in one or other spatial location. This means that PQT can predict the outcome of measurements of all the observables of OQT.

Fourth, insofar as the predictions of OQT and PQT differ, the difference is extraordinarily difficult to detect and will not be detectable in any quantum measurement so far performed.

**14.7 Crucial Experiments**

In principle, however, OQT and PQT yield predictions that differ for experiments that are extraordinarily difficult to perform and which have not yet, to my knowledge, been performed. Consider the following evolution:

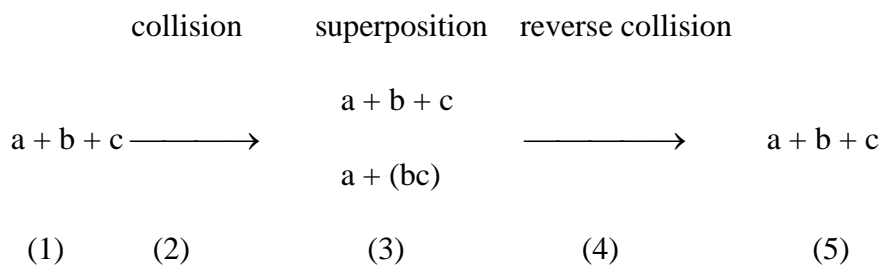

|  | collision | superposition | reverse collision |  |
|---|---|---|---|---|
|  |  | a + b + c |  |  |
| a + b + c | ⟶ |  | ⟶ | a + b + c |
|  |  | a + (bc) |  |  |
| (1) | (2) | (3) | (4) | (5) |

Suppose the experimental arrangement is such that, if the superposition at stage (3) persists, then interference effects will be detected at stage (5). Suppose, now, that at stage (3),

---

(II), modified in this way, would certainly procure sufficient localization to ensure that the classical world does not gradually become increasingly unlocalized.

[16] In fact, from a formal point of view (ignoring questions of interpretation), PQT has exactly the same structure as OQT with just one crucial difference: the generalized Born postulate of OQT is replaced by postulate (II) of section 5. (The generalized Born postulate specifies how probabilistic information about the results of measurement is to be extracted from the $\psi$-function.)

the condition for the superposition to collapse into one or other state, according to PQT, obtains. In these circumstances, OQT predicts interference at stage (5), whereas PQT predicts no interference at stage (5) (assuming this evolution is repeated many times). PQT predicts that in each individual case, at stage (3), the superposition collapses probabilistically into one or other state. Hence there can be no interference.

OQT and PQT make different predictions for decaying systems. Consider a nucleus that decays by emitting an $\alpha$-particle. OQT predicts that the decaying system goes into a superposition of the decayed and undecayed state until a measurement is performed, and the system is found either not to have decayed or to have decayed. PQT, in appropriate circumstances, predicts a rather different mode of decay. The nucleus goes into a superposition of decayed and undecayed states, which persists for a time until, spontaneously and probabilistically, in accordance with the postulate (II), the superposition jumps into the undecayed or decayed state entirely independent of measurement. The decaying system will continue to jump, spontaneously and probabilistically, into the undecayed state until, eventually, it decays.

These two processes of decay are, on the face of it, very different. There is, however, just one circumstance in which these two processes yield the same answer, namely if the rate of decay is exponential. Unfortunately, the rate of decay of decaying systems, according to quantum theory, *is* exponential. It almost looks as if nature is here maliciously concealing the mode of her operations. It turns out, however, that for long times, quantum theory predicts departure from exponential decay (Fonda et al., 1978). This provides the means for a crucial experiment. OQT predicts that such long-time departure from exponential decay will, in appropriate circumstances, obtain, while PQT predicts that there will be no such departure. The experiment is, however, very difficult to perform because it requires that the environment does not detect or "measure" decay products during the decay process. For further suggestions for crucial experiments, see Maxwell (1988, pp. 37–38).

There is a sense, it must be admitted, in which PQT is not falsifiable in these crucial experiments. If OQT is corroborated and PQT seems falsified, the latter can always be salvaged by letting $\varepsilon$, the undetermined constant of PQT, be sufficiently minute. Experiments that confirm OQT only set an upper limit to $\varepsilon$. There is always the possibility, however, that OQT will be refuted and PQT will be confirmed.

It would be interesting to know what limit present experiments place on the upper bound of $\varepsilon$.

### 14.8 The Potential Achievements of PQT

PQT provides a very natural possible solution to the quantum wave/particle dilemma. The theory is fully micro-realistic; it is, in the first instance, exclusively about "beables" to use John Bell's term. It makes sense of the mysterious quantum world. There is no reference to observables, to measurement, to macroscopic, quasi-classical, or irreversible phenomena or processes, or to the environment, whatsoever. As a result, PQT does not suffer from the eight defects, indicated in section 14.2, which beset OQT. The theory is restricted, in the first instance, to specifying how quantum micro-systems—quantum propensitons—evolve and interact with one another deterministically and probabilistically. But despite eschewing all reference to observables or measurement in its basic postulates, the theory nevertheless in principle recovers all the empirical success of OQT. At the same time, it is empirically distinct from OQT for experiments not yet performed and difficult to perform.

It is quite possible, of course, that the general propensiton solution to the wave/particle problem that I have outlined here is correct but the specific proposal that probabilistic transitions are to be associated with inelastic interactions is false. The falsity of the specific proposal does not mean that the general propensiton idea is false as well.

My chief objective in developing the version of PQT outlined here, from 1972 onwards, was to provoke physicists better qualified than I *to tackle the problem by doing straightforward physics*! I hoped physicists would put forward testable conjectures concerning the quantum conditions for probabilistic transitions to occur and then put these conjectures to the test of experiment. Straightforward physics, theoretical and experimental, is needed if we are to develop a more adequate, intelligible, and explanatory version of quantum theory than that which we possess today. This research is now, it seems, underway: see Bassi (2013).